\documentclass{article}
\usepackage{ijcai24}

\usepackage{times}
\usepackage{soul}
\usepackage{url}
\usepackage[hidelinks]{hyperref}
\usepackage[utf8]{inputenc}
\usepackage[small]{caption}
\usepackage{graphicx}
\usepackage{amsmath}
\usepackage{amsthm}
\usepackage{booktabs}
\usepackage{algorithm}
\usepackage{algorithmic}
\usepackage[switch]{lineno}
\usepackage{xcolor}
\usepackage{latexsym}
\usepackage{multirow}
\usepackage{colortbl}
\usepackage{array}
\usepackage{bm}

\title{EAT: Self-Supervised Pre-Training with Efficient Audio Transformer}

\author{
    \affiliations
    \emails
}

\author{
Wenxi Chen\and
Yuzhe Liang\and
Ziyang Ma\and
Zhisheng Zheng\and
Xie Chen \thanks{Corresponding author.}
\affiliations
MoE Key Lab of Artificial Intelligence, AI Institute, \\
X-LANCE Lab, Department of Computer Science and Engineering, \\ 
Shanghai Jiao Tong University, Shanghai, China \\
\emails
\{1029713857, chenxie95\}@sjtu.edu.cn
}
\begin{document}

\maketitle

\begin{abstract}

Audio self-supervised learning (SSL) pre-training, which aims to learn good representations from unlabeled audio, has made remarkable progress. 
However, the extensive computational demands during pre-training pose a significant barrier to the potential application and optimization of audio SSL models. 
In this paper, inspired by the success of data2vec 2.0 in image modality and Audio-MAE in audio modality, we introduce \textbf{E}fficient \textbf{A}udio \textbf{T}ransformer (EAT) to further improve the effectiveness and efficiency in audio SSL.
The proposed EAT adopts the bootstrap self-supervised training paradigm to the audio domain. A novel Utterance-Frame Objective (UFO) is designed to enhance the modeling capability of acoustic events. 
Furthermore, we reveal that the masking strategy is critical in audio SSL pre-training, and superior audio representations can be obtained with large inverse block masks. 
Experiment results demonstrate that EAT achieves state-of-the-art (SOTA) performance on a range of audio-related tasks, including AudioSet (AS-2M, AS-20K), ESC-50, and SPC-2, along with a significant pre-training speedup up to $\sim$15x compared to existing audio SSL models.~\footnote{The code and pre-trained models will be available at \\ \url{https://github.com/cwx-worst-one/EAT}. }  
% \url{https://github.com/cwx-worst-one/EAT}
% \url{https://github.com/anonymous}

\end{abstract}

\section{Introduction}

Self-supervised learning (SSL) has emerged as a pivotal method in audio representation learning, drawing inspiration from its success in natural language processing~\cite{devlin2018bert,radford2018improving}, computer vision~\cite{chen2020simple,he2020momentum}, and speech processing~\cite{hsu2021hubert,chen2022wavlm,ma2022mt4ssl}. 
The strength of SSL lies in leveraging vast amounts of unlabeled data, thus enabling models to effectively learn data features.
 % specific to their modality

Key to the success of SSL in the audio domain is masked autoencoder models and the bootstrap approach, celebrated for their ability to extract fruitful features from input data.
Reconstruction-based methods like BERT \cite{devlin2018bert}  and MAE \cite{he2021masked} learn representations by predicting global information from limited unmasked contexts. In contrast, BYOL \cite{grill2020bootstrap} and its derivatives implement data-augmentation-based prediction tasks for continuous self-learning with online and target networks.
Similar techniques have been adapted to develop audio SSL models. Models like SSAST \cite{gong2022ssast}, MAE-AST \cite{baade2022mae}, and Audio-MAE \cite{huang2022amae} concentrate on reconstructing audio spectrograms from masked patches. Others like BYOL-A \cite{niizumi2021byol}, ATST \cite{li2022atst}, and M2D \cite{niizumi2023masked} employ self-learning based on the bootstrap framework in augmented spectrogram data to learn latent audio representations during pre-training.

Despite these developments, the expensive computational cost of pre-training remains a hurdle. 
Approaches like Audio-MAE attempt to enhance encoding efficiency by using a high mask ratio and feeding only unmasked patches to the encoder. However, this would necessitate a complex decoder like a SwinTransformer~\cite{liu2021swin}, often leading to prolonged processes.  
Other audio SSL models aim to streamline pre-training by simplifying learning tasks. For example, in BEATs \cite{chen2022hts}, using a tokenizer to discretize target features allows it to emphasize semantically rich audio tokens and thereby facilitate learning in each iteration. However, this quantitative approach may result in the loss of objective information and require more pre-training iterations. 
% Lastly, the total training time remains relatively high, despite efforts to expedite the process.

Therefore, we introduce the \textbf{E}fficient \textbf{A}udio \textbf{T}ransformer (EAT) model, innovatively tailored for efficient learning of audio semantics and exceptional performance in downstream tasks. 
EAT departs from conventional methods that focus on reconstructing audio patches or predicting discrete features. Instead, it employs a unique Utterance-Frame Objective (UFO) during pre-training, synergizing global utterance-level and local frame-level representations in its prediction task. 
This dual-level objective incorporating global and local information from the audio spectrogram enhances the model's ability to understand audio clips.

As depicted in Figure \ref{fig: EAT model}, EAT employs a bootstrapping framework. The student model is continuously updated using target features from a teacher model, which in turn is progressively updated via an exponential moving average (EMA) technique, akin to MOCO \cite{he2020momentum}.

For the pretext task, EAT employs the Masked Language Modeling (MLM) with an 80\% masking ratio, focusing on patch embeddings from downsampled audio spectrograms accompanied by fixed sinusoidal positional embeddings.  
Inspired by the masking method in data2vec 2.0 \cite{baevski2023efficient} on image modality, EAT adopts an inverse block multi-mask technique on audio patches. This method preserves unmasked data in block units, resulting in larger regions of locality for unmasked patch embeddings and thus increasing the challenge of extracting audio semantics and predicting masked features. 
Additionally, the multi-mask strategy compensates for the computational cost associated with encoding the complete raw audio patches input to the teacher model during pre-training. Implementing multiple clones of masked input data (only masked parts encoded) in the student model significantly boosts data utilization efficiency.

At last, we design an asymmetric network architecture that combines a complex Transformer encoder with a lightweight CNN decoder. This setup efficiently decodes features, facilitating precise frame-level feature prediction.

With its efficient self-learning mechanism, the EAT model can adeptly acquire crucial audio features. Our experiments confirm that EAT, with significantly reduced training hours in total, achieves state-of-the-art performance on several audio and speech classification datasets, underscoring its superior generalization and learning efficiency in the audio domain.

Our contributions are summarized as follows: 
\begin{itemize}
\item We introduce a novel Utterance-Frame Objective (UFO) during pre-training in audio SSL for learning audio latent representation. The utterance-level learning is experimented to be crucial in model pre-training. 
\item  We adopt the inverse block multi-mask method from data2vec 2.0 with a high mask ratio on audio patches, which significantly speeds up the pre-training process in the audio bootstrap framework. Experiments show that EAT substantially outperforms previous audio SSL models in pre-training efficiency.
\item  We achieve SOTA results on several popular audio-related datasets.  The code and pre-trained models are also open-sourced to facilitate the development of the community.
\end{itemize}

% TODO:IJCAI记得去掉
\vspace{-0.2cm}
\section{Related Work}

\subsection{Bootstrap Method}
The concept of the bootstrap method was initially introduced in the context of self-supervised learning by BYOL \cite{grill2020bootstrap}. The BYOL architecture incorporates a dual-component framework, consisting of a target encoder and a predictor network. The target encoder is responsible for generating representative targets, while the predictor network aims to predict these targets using an augmented version of the input. The predictor network is updated through the prediction objective, whereas the target encoder undergoes momentum updates, a concept derived from the Momentum Contrast (MoCo) method \cite{he2020momentum}. This approach has inspired a series of subsequent self-supervised vision models, notable examples being DINO \cite{caron2021emerging}, SimSiam \cite{chen2021exploring}, and MoCo v3 \cite{chen2021empirical}.

Extending the bootstrap method to various modalities, data2vec \cite{baevski2022data2vec} and its successor, data2vec 2.0 \cite{baevski2023efficient}, represent significant advancements in self-supervised learning. These models utilize mask-based techniques for contrasting the pretext task, significantly enhancing pre-training efficiency. Their approach also involves regressing representations across multiple neural network layers, rather than concentrating exclusively on the top layer. 

In an endeavor to embrace the potential of the bootstrap method like BYOL-A \cite{niizumi2021byol} and M2D \cite{niizumi2023masked}, our EAT model also applies this methodology to the audio domain and aims to enhance the audio feature learning while improving the pre-training efficiency.

\subsection{Self-supervised Audio Pre-training}
Self-supervised learning (SSL) in the audio domain involves extensive pre-training using large volumes of unlabeled data to learn latent audio features. Typically, there are two main approaches to selecting in-domain pre-training data. The first approach is joint pre-training, which combines speech and audio data, as exemplified by models like SS-AST \cite{gong2022ssast} and MAE-AST \cite{baade2022mae}. The second, and more prevalent approach, is to exclusively use audio data for pre-training, as seen in models such as MaskSpec \cite{chong2023masked}, MSM-MAE \cite{niizumi2022masked}, Audio-MAE \cite{huang2022amae}, and our EAT model.

Various methods are employed in different components of audio SSL models. For input data, models like wav2vec 2.0 \cite{baevski2020wav2vec} and data2vec process raw waveforms, whereas most others including EAT use Mel spectrograms to extract features. In terms of pretext tasks, models employing Masked Language Modeling (MLM) techniques, such as MAE-AST,  Audio-MAE, and our EAT model, apply higher masking rates to audio patches. Contrastingly, models like BYOL-A \cite{niizumi2021byol} and ATST \cite{li2022atst} use augmentation techniques like mixup and random resize crop (RRC) to provide varied auditory perspectives.

The pre-training objectives also vary across models. For instance, Audio-MAE and MAE-AST use an MAE-style task, reconstructing original spectrogram patches where unmasked data predicts the masked ones. BEATs \cite{chen2022beats} employs a tokenizer for discretized semantic feature prediction. Meanwhile, models like data2vec, BYOL-A, and M2D, focus on predicting latent representations. In EAT, we have adapted the representation prediction task into the Utterance-Frame Objective (UFO) to take both global and local information in the audio spectrogram into consideration.

\begin{figure*}[t]
  \centering
  \includegraphics[width=1\textwidth]{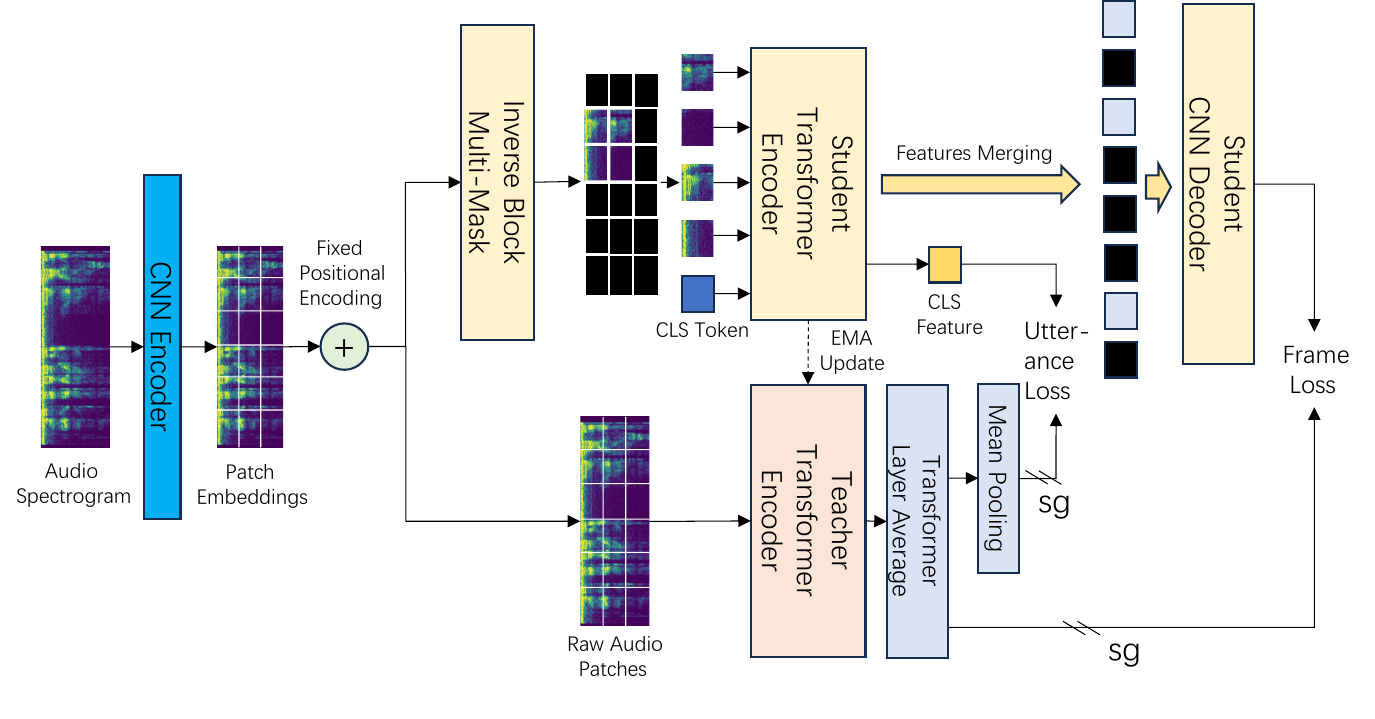}
  \caption{\textbf{Architecture of EAT in Audio Self-supervised Pre-training.} EAT first transforms the audio spectrogram into patch embeddings with a CNN encoder. They are then separately fed into the student model via the inverse block multi-mask method and the teacher model with the same network directly.  Subsequently, the generated features merged with the masked parts, are decoded using a lightweight CNN decoder. The teacher model synthesizes the average output from all Transformer layers as the target value. The utterance-level loss utilizes regression on the mean pooling values of the target values across patch dimensions, while the frame-level loss uses regression on target values at masked positions. The teacher model is updated through the EMA method, based on the learnable parameters of the student model. Notably, ``sg'' means stop-gradient here.  }
  \label{fig: EAT model}
\end{figure*}

\section{Method}

EAT draws inspiration from the data2vec 2.0 \cite{baevski2023efficient} and Audio-MAE \cite{huang2022amae} model, incorporating a blend of bootstrap and masked modeling method to effectively learn the latent representations of audio spectrogram. In this process, we devised an asymmetric network architecture that employs a standard Transformer encoder for processing visible patches (unmasked regions) and a lightweight CNN decoder for the comprehensive decoding of all features, including those at masked positions. This architecture enables rapid pre-training: complex encoding is applied to smaller data (visible patches), while a simpler decoder processes the entire data (visible features along with masked tokens). Furthermore, EAT distinctively combines frame-level loss, focusing on latent representation reconstruction, with utterance-level loss, targeting global representation prediction. This simple combination allows the model to adeptly capture both local nuances and overarching trends from raw audio data, significantly enhancing its performance. Figure \ref{fig: EAT model} illustrates our EAT model and the details of each component, pre-training and fine-tuning are as follows. 

\subsection{Model Architecture}

% \subsubsection{Patch Embedding with Positional Encoding}
\paragraph{Patch Embedding with Positional Encoding.} EAT is designed to operate on audio spectrograms rather than the original waveforms. To downsample audio spectrogram features, we first use padding to extend it along the time frame to a uniform length (suitable for different datasets), and then extract patch embeddings from it through a 2D convolutional layer encoder. We maintain the CNN encoder's kernel size $S$ and stride the same to ensure the relative independence between patch embeddings by preventing overlap. Specifically, the audio spectrogram $\textbf{X} \in R^{T\times F}$ is transformed into patch embeddings $\textbf{X}_p \in R^{P\times E}$, where $T\times F$ represents the time and frequency dimensions of the input spectrogram and $P\times E$ denotes the patch size and embedding features dimensions, with $P = TF / S^2 $ after flattening. 
Subsequently, 1D fixed positional encoding used in standard ViT \cite{dosovitskiy2020image} is applied to these embeddings, providing essential positional information for more effective encoding in subsequent Transformer blocks.

\subsubsection{Utterance-Frame Objective}
EAT introduces a Utterance-Frame Objective (UFO)  function during pre-training, effectively merging global utterance-level and local frame-level losses in audio representation prediction. This dual-focus strategy is a significant advancement in contextualized target prediction.

The contextualized target $\textbf{Y}_a \in {R}^{P \times E}$ is derived from the top-k-layers of the Transformer blocks output in the teacher model, processing complete input patch embeddings $\textbf{Y}_r \in {R}^{P \times E}$. Unlike the BYOL \cite{grill2020bootstrap} method, which utilizes only the last layer's output feature as target, EAT computes $\textbf{Y}_a$ by averaging outputs across all Transformer layers. This approach ensures a comprehensive representation target that captures both shallow-level, raw audio features and deep-level, semantically rich latent representations.

To effectively integrate global utterance information from audio spectrograms without adding structural complexity, EAT incorporates a simple, learnable classification token (CLS token) into the student model. The multi-head self-attention mechanism of the Transformer architecture allows this CLS token $\textbf{c} \in R^{1\times E}$ to view and access information from all unmasked patch embeddings. Then, we use the CLS feature $\textbf{c}'\in {R}^{1 \times E}$ from student encoder output to predict the average value of $\textbf{Y}_a$ in patch dimension, i.e. $\textbf{y}_a' \in {R}^{1 \times E}$, with MSE loss. The utterance loss is calculated as follows: 

$$
L_u = || \textbf{c}' -  \textbf{y}_a'||^2_2
$$

Distinctively, EAT's approach to utterance-level learning sets it apart from models like ATST-Clip \cite{li2023selfsupervised}. EAT avoids additional projectors or predictors for feature transformation, directly focusing on capturing global audio features at the utterance level. This direct regression technique is experimentally shown to effectively preserve crucial information in global audio representation learning, reducing the risk of information loss during feature transformation.

For local frame-level learning in the audio patches, EAT employs the MAE \cite{he2021masked} method. The student encoder output representations $\textbf{X}_d \in {R}^{P'\times E}$, merged with mask tokens from the original sequence, predict the average features $\textbf{Y}_a$ at masked positions using a lightweight CNN decoder. The frame loss, also based on MSE, estimates the difference between the decoder output $\textbf{X}_o \in {R}^{P''\times E}$ and the target value $\textbf{Y}_o \in {R}^{P''\times E}$, where $P'' = T' \times F' \times M$. The frame loss is computed as:

$$
L_f = || \textbf{X}_o  -  \textbf{Y}_o ||^2_2
$$

Finally, the UFO loss by combining the frame-level and utterance-level losses can be given by:

$$
L_{UFO} = L_f + \lambda L_u 
$$

$\lambda$ is the hyperparameter to determine the impact of utterance loss and is found to be crucial to the overall performance of EAT as shown in Section \ref{mask experiment}.

\subsubsection{Masking Strategies in Pre-training}
A pivotal element contributing to the EAT model's efficiency in learning audio representations is the masking strategy.
In our EAT model, a masking rate of up to 80\% is employed for patch embeddings before encoding. This high masking rate substantially reduces the data volume processed by the Transformer, akin to the approach in MAE, thereby enhancing training speed. More importantly, it escalates the challenge of masked learning, compelling the model to decipher the essential information from the entire audio spectrogram with more limited visible input and infer the masked features during pre-training.

\begin{figure}[htbp]
  \centering
  \includegraphics[width=0.48\textwidth]{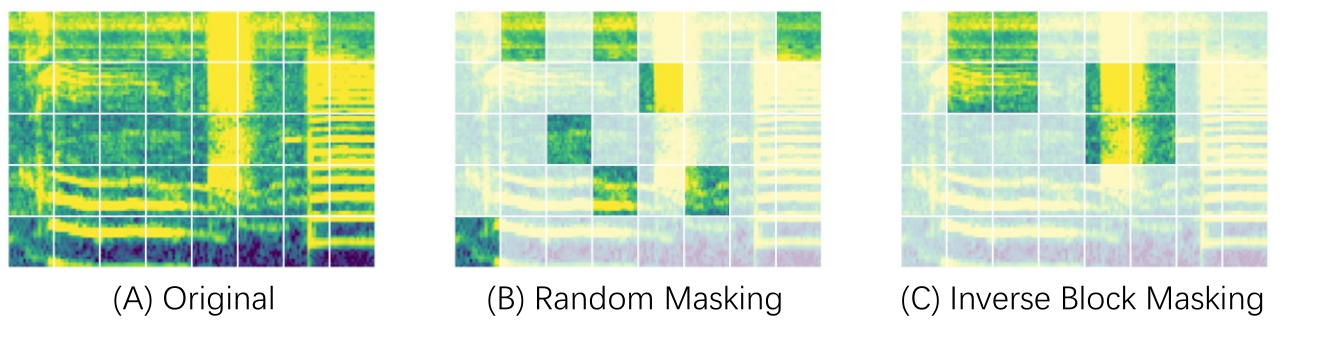}
  \caption{\textbf{Inverse Block Masking on Audio Patches.} The block size is set to $2 \times 2$ with a masking ratio of $80\%$ in the right subfigure.  }
  \label{fig: inverse block mask}
\end{figure}

The masking method of EAT, as depicted in Figure \ref{fig: inverse block mask}, is distinct from previous audio SSL models. Instead of random masking audio patches, EAT implements inverse block masking proposed in data2vec 2.0 \cite{baevski2023efficient}  on image modality. 
For a given patch embedding $\textbf{X}_p \in {R}^{P \times E}$, instead of applying 1D random masking which decorrelates the time and frequency dimensions, EAT's masking reshapes $\textbf{X}_p$ into $\textbf{X}_p' \in {R}^{T'\times F'\times E}$ and applies a 2D random mask. This mask maintains correlation in both time and frequency dimensions, where $T' = T / S$ and $F' = F / S$. 
The process involves initially masking all patches, then iteratively preserving original parts in block size until the masked embedding count aligns with the desired masking rate. Compared to 1D random masking with the same masking ratio, it challenges EAT to concentrate on a more restricted yet focused set of fragmented audio clips for representation prediction using UFO.
% This approach decreases the mutual information between visible parts and masked ones in the patch embeddings, making it more challenging for the student model to predict the latent representation in both the global utterance and the local frames. 

In addition, EAT could be further accelerated with the multi-mask approach. The teacher model, processing complete patch embeddings, demands greater computational resources for encoding than its student.
% It would be inefficient to use only one corresponding audio patch as input to the student model at a time, especially in our experiment with a relatively high mask ratio. 
To optimize efficiency, EAT employs the multi-mask strategy, creating multiple clone-masked embeddings from the same spectrogram patch using different inverse block masking. These variants are concurrently inputted into the student model, thus amplifying data utilization via parallel computing.

\subsection{Pre-training Details}

The EAT model consists of 93M parameters during pre-training and 88M in fine-tuning (post-CNN decoder released), aligning with the parameter scale of other standard base audio SSL models. We employ a CNN encoder with a (16,16) kernel and a stride of 16 for downsampling audio spectrograms, ensuring non-overlapping patch features extraction in the time and frequency dimensions. Both student and teacher model encoders use the 12-layer ViT-B \cite{dosovitskiy2020image} model. For faster decoding, EAT utilizes a 6-layer 2D CNN decoder with (3,3) kernels, LayerNorm \cite{ba2016layer}, and GELU activation \cite{hendrycks2016gaussian}
. 
% group-parameterized \cite{krizhevsky2012imagenet} for efficiency.

During the self-supervised pre-training, the student model with parameters $\theta_s$ is updated via the UFO function. Following the general bootstrap approach, the teacher model with parameters $\theta_t$ in EAT is updated using an Exponential Moving Average (EMA) strategy. The parameter update formula \cite{lillicrap2015continuous} is defined as:

$$
\theta_t \leftarrow \tau \theta_t + (1-\tau) \theta_s
$$

EAT employs a linearly increasing strategy for adjusting the value of $\tau$. This approach provides the model with enhanced flexibility and randomness in the initial training stages, facilitating parameter adjustments and supporting the learning process of the student model. As training advances, $\tau$ approaches 1, leading to a more stable learning.

\subsection{Fine-tuning Details}

In the fine-tuning stage, EAT generates latent representations using the student Transformer encoder and replaces the original CNN student decoder with a linear layer for predicting audio categories. Additionally, we implement several data augmentation techniques to fully exploit the model's acquired comprehension of audio spectrogram features from the pre-training phase. It is crucial in enhancing EAT's domain adaptation capabilities for specific downstream tasks.

During fine-tuning, EAT is enhanced with audio augmentations including SpecAug \cite{park2019specaugment}, mixup \cite{zhang2017mixup}, droppath \cite{huang2016deep}, audio rolling, and random noise. Specifically, mixup is applied to spectrograms, aligning with EAT's pre-training focus on spectrogram-based latent representations. 
For classification tasks, a CLS token is used for final prediction, which shows improved performance over mean pooling methods in our experiments in Section \ref{utterance learning}.

\begin{table*}[t]
\centering
\begin{tabular}{lcccccc}
\hline
\multirow{2}{*}{\textbf{Model}} & \multirow{2}{*}{\#Param} & {Pre-training} & AS-2M & AS-20K & ESC-50 & SPC-2 \\
& &  Data & mAP(\%) & mAP(\%)  & Acc(\%)   & Acc(\%) \\
\hline
\textbf{Supervised Pre-Training} \\ 
\textcolor{gray!50}{PANN}  \cite{kong2020panns} & \textcolor{gray!50}{81M} & \textcolor{gray!50}{-} & \textcolor{gray!50}{43.1} & \textcolor{gray!50}{27.8} & \textcolor{gray!50}{83.3}  & \textcolor{gray!50}{61.8}\\ 
\textcolor{gray!50}{PSLA} \cite{gong2021psla} & \textcolor{gray!50}{14M} & \textcolor{gray!50}{IN} & \textcolor{gray!50}{44.4} & \textcolor{gray!50}{31.9}  & \textcolor{gray!50}{-}  & \textcolor{gray!50}{96.3} \\
\textcolor{gray!50}{AST} \cite{gong2021ast} & \textcolor{gray!50}{86M}& \textcolor{gray!50}{IN} & \textcolor{gray!50}{45.9} & \textcolor{gray!50}{34.7} & \textcolor{gray!50}{88.7} & \textcolor{gray!50}{98.1} \\
\textcolor{gray!50}{MBT} \cite{nagrani2021attention} & \textcolor{gray!50}{86M} & \textcolor{gray!50}{IN-21K} & \textcolor{gray!50}{44.3} & \textcolor{gray!50}{31.3} & \textcolor{gray!50}{-} & \textcolor{gray!50}{-}\\
\textcolor{gray!50}{PassT} \cite{koutini2021efficient} & \textcolor{gray!50}{86M}& \textcolor{gray!50}{IN} & \textcolor{gray!50}{47.1} & \textcolor{gray!50}{-} & \textcolor{gray!50}{96.8} & \textcolor{gray!50}{-}  \\
\textcolor{gray!50}{HTS-AT} \cite{chen2022hts} & \textcolor{gray!50}{31M}& \textcolor{gray!50}{IN} & \textcolor{gray!50}{47.1} & \textcolor{gray!50}{-}  & \textcolor{gray!50}{97.0}   & \textcolor{gray!50}{98.0}\\
\textcolor{gray!50}{Wav2CLIP} \cite{wu2022wav2clip} & \textcolor{gray!50}{74M}& \textcolor{gray!50}{TI+AS} & \textcolor{gray!50}{-} &  \textcolor{gray!50}{-} &  \textcolor{gray!50}{86.0} & \textcolor{gray!50}{-} \\
\textcolor{gray!50}{AudioCLIP} \cite{guzhov2022audioclip} & \textcolor{gray!50}{93M}& \textcolor{gray!50}{TI+AS} & \textcolor{gray!50}{25.9} & \textcolor{gray!50}{-} & \textcolor{gray!50}{96.7} & \textcolor{gray!50}{-}\\

\hline
\textbf{Self-Supervised Pre-Training}\\
% wav2vec 2.0 \cite{baevski2020wav2vec}  & 95M & LS & - & - & - & - \\
Conformer \cite{srivastava2022conformer} & 88M & AS & 41.1 & - & 88.0 & -\\
SS-AST \cite{gong2022ssast} & 89M & AS+LS & - & 31.0 & 88.8 & 98.0\\
MAE-AST \cite{baade2022mae}& 86M & AS+LS & - & 30.6 & 90.0  & 97.9 \\
MaskSpec \cite{chong2023masked}& 86M & AS & 47.1 & 32.3 & 89.6  & 97.7 \\
MSM-MAE \cite{niizumi2022masked}& 86M & AS & - & -  & 85.6 & 87.3 \\
data2vec \cite{baevski2022data2vec}& 94M & AS & - & 34.5 & -  & -\\
Audio-MAE \cite{huang2022amae}& 86M & AS & 47.3 & 37.1 & 94.1 & \textbf{98.3} \\
$\text{BEATs}_{iter1}$ \cite{chen2022beats} & 90M & AS & 47.9 & 36.0 & 94.0 & \textbf{98.3} \\
$\text{BEATs}_{iter2}$ \cite{chen2022beats} & 90M & AS & 48.1 & 38.3 & 95.1 & \textbf{98.3} \\
$\text{BEATs}_{iter3}$ \cite{chen2022beats} & 90M & AS & 48.0 & 38.3 & 95.6 & \textbf{98.3} \\
\textcolor{gray!50}{$\text{BEATs}_{iter3+}$}  \cite{chen2022beats} $^*$ & \textcolor{gray!50}{90M} & \textcolor{gray!50}{AS} & \textcolor{gray!50}{48.6}  & \textcolor{gray!50}{38.9} & \textcolor{gray!50}{98.1} & \textcolor{gray!50}{98.1}\\
\hline

\textbf{Ours}\\
EAT & 88M & AS & \textbf{48.6} & \textbf{40.2} & \textbf{95.9} & \textbf{98.3} \\
\hline

\end{tabular}
\caption{\textbf{Model Comparison among existing methods in audio classification tasks.} Pre-training data sources include ImageNet (IN), AudioSet (AS), and LibriSpeech (LS), while CLIP utilizes 400M text-image pairs (TI). We  \textcolor{gray!50}{gray-out} the methods with additional supervised training on external datasets or additional pseudo-labels. $^*$: Models employ knowledge distillation across iterations with extra pseudo-labels.  }
\label{sota model comparison}
\end{table*}

 % Methods supplemented by external dataset training or pseudo-labels are \textcolor{gray!50}{grayed-out}. }

\section{Experiments}
We pre-trained EAT on the AudioSet-2M (AS-2M) dataset \cite{gemmeke2017audio}, evaluating its performance through audio-classification fine-tuning on AS-2M, AS-20K, and the Environmental Sound Classification (ESC-50) \cite{piczak2015esc} datasets, as well as speech-classification fine-tuning on the Speech Commands V2 (SPC-2) \cite{warden2018speech} dataset.

\subsection{Experimental Setups}

% \subsubsection{AudioSet (AS-2M, AS-20K)}
\paragraph{AudioSet (AS-2M, AS-20K).} AudioSet, comprising approximately two million YouTube video audio clips of 10 seconds each, spans 527 classes. In our experiment, we downloaded and processed 1,912,134 clips as the unbalanced set (AS-2M) and 20,550 as the balanced set (AS-20K), with an evaluation set of 18,884 clips. Given the multi-category nature of these clips, we employed mean Average Precision (mAP) as our test metric, which calculates the average precision across multiple classes.

% \subsubsection{Environmental Sound Classification (ESC-50)}
\paragraph{Environmental Sound Classification (ESC-50).} ESC-50 dataset consists of 2,000 audio clips, each five seconds long and distributed across 50 semantic classes. In our evaluation, we implemented a five-fold cross-validation method, using 400 clips for validation and the remaining for training in each fold. The evaluation metric is the average validation accuracy across five folds in audio classification. 
% Accuracy here is defined as the proportion of correctly predicted clips out of the total number of clips. 

\paragraph{Speech Commands V2 (SPC-2).} SPC-2 is a keyword-spotting task in speech recognition, comprising 35 specific speech commands. It includes 84,843 training recordings, 9,981 validation recordings, and 11,005 testing recordings, each lasting 1 second. We utilized the data split from the SUPERB \cite{yang2021superb} benchmark to evaluate accuracy.

\subsubsection{Training Details}

We uniformly resampled the input waveforms to 16kHz sample rate, then transformed them into 128-dimensional Mel-frequency bands using a 25ms Hanning window with a 10ms shift. To preserve edge features during feature extraction with the CNN encoder, padding was applied to the Mel spectrogram. Additionally, the audio spectrogram patches are then normalized with a mean value of 0 and a standard deviation of 0.5, following the approach used in previous works.

EAT was pre-trained using AS-2M for 10 epochs with a batch size of 12 and a peak learning rate of 0.0005. For each clip, we created 16 clones with different inverse block masks via the multi-mask method. The cosine annealing learning strategy with warm-up steps \cite{loshchilov2016sgdr} was employed, alongside the Adam optimizer \cite{loshchilov2017decoupled}, with $\beta_1$ and $\beta_2$ values set to 0.9 and 0.95, respectively. We distribute the training load over 4 RTX 3090 GPUs and the total training time is around 58 hours.

\subsection{Main Results}

\subsubsection{Model Performance}

Table \ref{sota model comparison} presents the classification evaluation results of EAT and other audio models on AS-2M, AS-20K, ESC-50, and SPC-2 datasets, respectively. We categorize them into Supervised Pre-Training and Self-supervised Pre-Training models. For fair comparison, our performance evaluation benchmark primarily focuses on Self-supervised Pre-Training models.

In the audio classification task, the EAT model achieved SOTA performance on AS-2M, AS-20K, and ESC-50 datasets. On the AS-2M dataset, EAT achieved a mAP evaluation of 48.6\%, outperforming the previous SOTA value by 0.6\%. On the AS-20K dataset, EAT reached an impressive mAP of 40.2\%, surpassing the previous SOTA by 1.9\%. Moreover, in the ESC-50 dataset, EAT demonstrated exceptional accuracy, achieving 95.9\%, effectively reducing the average error rate from 4.4\% to 4.1\%. These results underscore EAT's robust ability to capture and interpret both global and local features in audio data, leading to outstanding performance in these challenging audio classification tasks.

In the domain of speech classification, EAT demonstrated commendable performance as well. Although our primary experimental focus was on audio datasets, EAT's proficiency was equally evident in speech classification tasks, notably in SPC-2. Here, EAT attained competitive accuracies, reaching 98.3\%, which aligns with the performance of previous SOTA models. This outcome underscores EAT's versatility and its broad applicability across various audio and speech tasks.

\begin{table}[htbp]
\centering
\begin{tabular}{c c c c c}
\toprule
{model} & {epoch} &{hour $\times$ GPU} & {speedup} & mAP \\
\midrule
$\text{BEATs}_{iter3}$ & 342 & 3600 & $ 1 \times$ & 38.3 \\
Audio-MAE & 32 & 2304 & $ 1.56 \times$ & 37.1\\
{\textbf{EAT}}  & \textbf{10} & \textbf{ 230 } &\textbf{ 15.65$\times$}  & \textbf{40.2} \\
\bottomrule
\end{tabular}
\caption{ \textbf{Comparison with $\text{BEATs}_{iter3}$ and Audio-MAE on pre-training cost.} We evaluate the pre-training wall-clock time of EAT on 4 RTX 3090 GPUs in Fairseq \protect\cite{ott2019fairseq} and it demands around 5.8 hours for each epoch. BEATs is pre-trained on 16 Tesla V100-SXM2-32GB GPUs for around 75 hours per iteration with 114 epochs while Audio-MAE on 64 V100 GPUs for approximately 36 hours in total. All models are uniformly fine-tuned on AS-20K. }
\label{Efficiency comparison}
\end{table}

% 这个要重新测
\begin{figure}[htbp]
  \centering
  \includegraphics[width=0.48\textwidth]{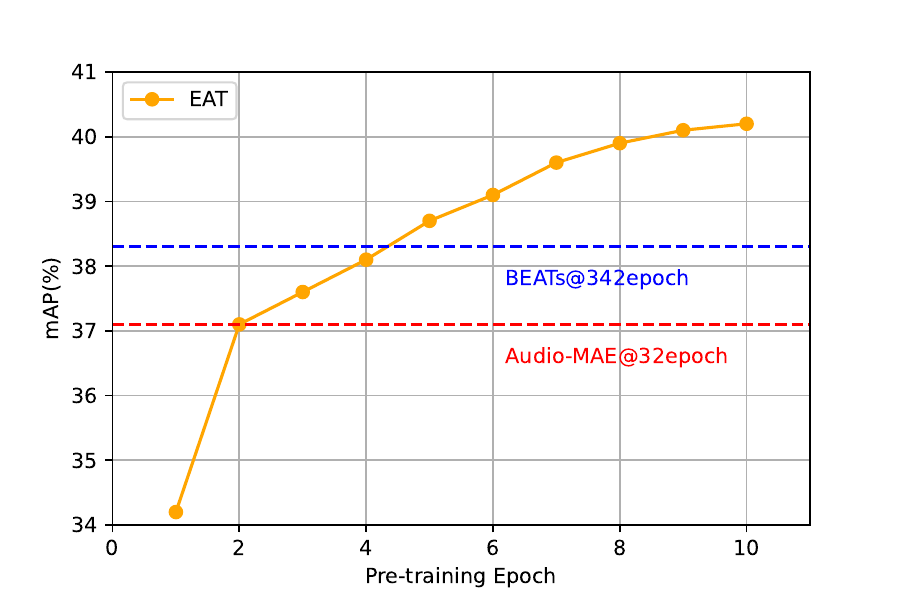}
  \caption{\textbf{Comparison with $\text{BEATs}_{iter3}$ and Audio-MAE on pre-training epoch during EAT's 10-epoch pre-training.} All models are 
uniformly fine-tuned on AS-20K and tested on the evaluation set.}
  \label{fig: Epoch comparison}
\end{figure}

\subsubsection{Pre-training Efficiency}

The EAT model showcases exceptional efficiency during its pre-training phase compared to previous SOTA audio self-supervised learning models. As depicted in Table \ref{Efficiency comparison}, EAT, pre-trained for just 10 epochs, achieves a total pre-training time reduction of 15.65 times compared to $\text{BEATs}_{iter3}$ and 10.02 times relative to Audio-MAE. 
Furthermore, as shown in Figure \ref{fig: Epoch comparison}, EAT matches Audio-MAE's performance after only two epochs and surpasses $\text{BEATs}_{iter3}$ by the fifth epoch. 
This substantial enhancement in training efficiency greatly reduces the computational resources, easing the pre-training process for a high-performing base audio SSL model.

The efficiency gains of EAT are attributable to two key aspects. 
First, EAT adopts a high mask ratio of 80\% during pre-training. This substantial masking implies that a significant portion of audio data is excluded before being fed to the student encoder, enhancing batch processing capacity and exploiting the benefits of parallel computing for improved efficiency. 
Second, our proposed Utterance-Frame Objective (UFO) function, a departure from the traditional audio spectrogram patch reconstruction objective, requires only lightweight decoding. Thus, EAT employs a lightweight CNN decoder for feature decoding and prediction, in contrast to the Transformer blocks used in models like Audio-MAE, which greatly speeds up the pre-training process.

Table \ref{Efficiency comparison} and Figure \ref{fig: Epoch comparison} illustrate EAT's efficiency in delivering outstanding performance with a small number of training epochs. Notably, EAT outperforms Audio-MAE (pre-trained for 32 epochs) and $\text{BEATs}_{iter3}$ (pre-trained for 342 epochs) on AS-20K after just 5 epochs of pre-training. Furthermore, after a total of 10 epochs of pre-training, EAT's mAP test score reached 40.2\%, markedly outperforming these two models that underwent more extensive pre-training.

This impressive performance with fewer training epochs is largely due to the EAT's multi-mask strategy on the audio spectrogram during the utterance-frame union pre-training. By employing multiple clones with different block masking on the audio patch embeddings, EAT effectively learns to ``listen'' to fragmented audio from various perspectives, enabling a more comprehensive understanding and superior performance than the single-angle one. Despite the reduction in input batch size per update, this strategy significantly enhances the data utilization of each audio clip, thereby substantially improving the efficiency of the EAT model.

\subsection{Ablation Study}

We conduct comprehensive ablation studies to evaluate the contributions of key components in EAT.  These studies evaluated different configurations of EAT, all pre-trained for 10 epochs on AS-2M and then fine-tuned on AS-20K.

\subsubsection{Utterance-level Learning\label{utterance learning}}

\begin{figure}[htbp]
  \centering
  \includegraphics[width=0.48\textwidth]{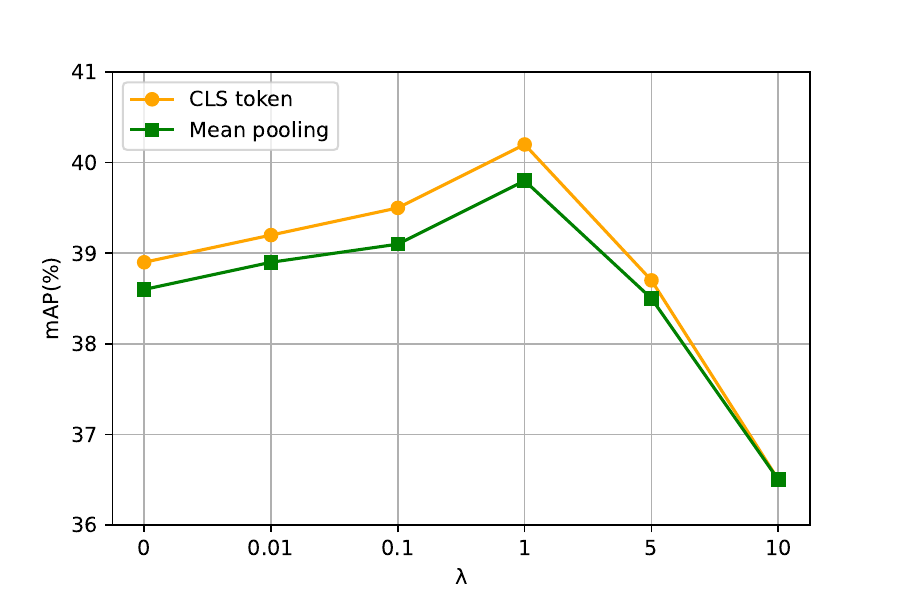}
  \caption{\textbf{Comparison on Utterance-level Loss Weight $\lambda$ in Pre-training and Prediction Methods in Fine-tuning.} During fine-tuning, we compare the effect of the final prediction on using the CLS token and mean pooling over all frames.}
  \label{fig:cls comparison}
\end{figure}

Our experiments delved into the significance of utterance-level learning by analyzing the impact of the utterance loss weight $\lambda$ during pre-training, as well as the effectiveness of the CLS-token-predicting method during fine-tuning.

Figure \ref{fig:cls comparison} illustrates that incorporating utterance loss $L_u$ alongside frame loss $L_f$ notably enhances the performance of EAT. Adopting a balanced approach with an utterance to frame loss weight ratio of 1:1 ($\lambda = 1$) not only provides a 1.3\% increase in mAP over a model configuration with no utterance loss ($\lambda = 0$) but also shows a 1.0\% improvement compared to a skewed ratio of 1:100 ($\lambda = 0.01$). However, an excessively high utterance loss weight ($\lambda = 10$) results in diminished performance, indicating that overemphasis on utterance-level learning can compromise the model's overall understanding abilities on audio clips.

Additionally, as Figure \ref{fig:cls comparison} shows, our experiment reveals a distinct advantage in using the CLS token for predictions over the mean pooling method. While mean pooling, averaging encoder output features across the patch dimension, is commonly effective in many audio SSL models, EAT's focus on global features through increased utterance loss weight during pre-training enhances the learnable CLS token's ability to extract global features. Consequently, this approach leads to improved performance of EAT in classification tasks.

In summary, appropriately weighting the utterance loss during pre-training enhances EAT's focus on global audio spectrogram features, fostering a more comprehensive latent representation learning. Additionally, using the CLS token for prediction in fine-tuning further boosts the model's performance, leveraging these global features for improved audio classification.

\subsubsection{Inverse Block Masking on Audio Patches\label{mask experiment}}

\begin{table}[htbp]
\centering
\begin{tabular}{c c}
\toprule
{Block Size} & {mAP(\%)} \\
\midrule
$1 \times 1 $ & 37.8 \\
$2 \times 2$ & 39.5 \\
$3 \times 3$ & 39.9 \\
$4 \times 4$  & 40.0 \\
$5 \times 5$ & \textbf{40.2}\\
$6 \times 6$ & 39.8 \\
$7 \times 7$ & 39.8 \\
$8 \times 8$ & 39.8 \\
\hline
$5 \times 5, 6 \times 4, 8 \times 3$ & \textbf{40.2} \\

\bottomrule
\end{tabular}
\caption{\textbf{Comparison on different block sizes during EAT pre-training within the inverse block masking on audio patches.}}
\label{Block length comparison}
\end{table}

In exploring the impact of the masking strategy during pre-training, we observed notable differences in EAT's performance. Table \ref{Block length comparison} illustrates that the inverse block masking (with block size $S > 1\times1 $) on audio patches performs better compared to the random masking ($S = 1 \times 1$). Notably, EAT configured with an increased inverse block size of $S = 5 \times 5$ attained the highest evaluation mAP of 40.2\%.

We conducted experiments with flexible block size sampling for masking, allowing the model to randomly preserve audio patches in block size like $5\times5$, $6\times4$, and $8\times3$ during pre-training. The outcomes were similar to using only $5\times5$ blocks, suggesting that block shape has a limited impact on performance. 
Instead, the key factors are block size and quantity in the mask. With a fixed 80\% mask ratio, properly increasing the block size (and correspondingly, reducing the total number of preserved blocks) in the mask is instrumental in enhancing the model's performance. 
When the block size is small, numerous preserved blocks scattered across audio patches make it easier for the model to deduce masked parts, limiting its ability to deeply understand audio representations. 
Conversely, using sufficiently large blocks for inverse masking effectively reduces the mutual information between visible and masked audio patches, aiding the model in learning to extract features from a more constrained set of known information and predict the unknown patches.

\section{Conclusion}

In this paper, we propose an  \textbf{E}fficient \textbf{A}udio \textbf{T}ransformer (EAT) model for effective and efficient audio-based self-supervised learning. EAT stands out by significantly expediting the pre-training process and delivering exceptional performance. Central to EAT's design is the novel use of the  Utterance-Frame Objective (UFO) loss, which is proven instrumental in learning audio latent representations. The integration of utterance-level learning, enhanced by balancing its loss weight with the frame-level learning during pre-training and employing CLS-token-based prediction in fine-tuning, effectively captures global audio features. EAT achieves state-of-the-art (SOTA) results in several audio and speech classification tasks, including AudioSet, ESC-50, and SPC-2, surpassing existing base audio SSL models in overall performance. The implementation of an inverse block multi-mask method with a high mask ratio on audio spectrogram patches contributes to EAT's expedited pre-training, outpacing models like Audio-MAE and BEATs by more than tenfold in terms of time efficiency.

In the future, we plan to scale up EAT to further explore its performance potential. Additionally, we aim to investigate audio-speech joint training, delving into the interplay between these two domains using our EAT model.

% \section*{Ethical Statement}

% There are no ethical issues.

\section*{Acknowledgments}
We thank Sanyuan Chen for the helpful discussions and feedback. 
% 盲审时不能放

\bibliographystyle{named}
\bibliography{ijcai24}

% \newpage
\appendix
\section{Appendix}
\subsection{Hyperparamter Settings}

Table \ref{tab:training_config} shows the hyperparameter settings for the pre-training and fine-tuning phases of EAT. For efficiency and to maintain a lightweight experimental setup, we uniformly utilized four GPUs during pre-training and employed a single GPU for fine-tuning.

\begin{table*}[t]
\centering
\begin{tabular}{l|c|cccc}
\toprule
\multirow{2}{*}{Hyperparameters} & Pre-Training & \multicolumn{4}{c}{Fine-Tuning}  \\
 & AS-2M & AS-2M & AS-20K & ESC-50 & SPC-2  \\
\midrule
Optimizer & \multicolumn{5}{c}{AdamW \cite{loshchilov2017decoupled}} \\
Optimizer Momentum &  \multicolumn{5}{c}{$\beta_1 = 0.9, \beta_2 = 0.95$} \\
Weight Decay &  \multicolumn{5}{c}{0.05} \\
Learning Rate Schedule & \multicolumn{5}{c}{Cosine \cite{loshchilov2016sgdr}} \\
Peak Learning Rate & 0.0005 & 0.00005 & 0.00005 & 0.00005 & 0.0002 \\
Minimum Learning Rate & \multicolumn{5}{c}{0.000001} \\
Steps & 400K & 300K & 40K & 4K & 40K  \\
Warm-up steps & 53K & 30K & 4K & 400 & 4K \\
Batch size & 12 & 96 & 48 & 48 & 256  \\
Clone batch & 16 & \multicolumn{4}{c}{N/A}  \\
GPUs & 4 & \multicolumn{4}{c}{1} \\
Dropout \cite{srivastava2014dropout} & 0.0 & 0.0 & 0.0 & 0.0 & 0.0  \\
Drop path \cite{huang2016deep} & 0.0 & 0.1 & 0.1 & 0.1 & 0.1  \\
Weighted Sampling & False & True & False & False & False  \\
Weighted Sampling size & N/A & 200K & N/A & N/A & N/A  \\
Roll Augmentation & False & True & True & True & False  \\
Noise Augmentation & False & False & False & False & True  \\
SpecAug \cite{park2019specaugment} &  N/A & 0.2 & 0.2 & 0.2 & 0.1  \\
Mixup \cite{zhang2017mixup} & 0.0 & 0.8 & 0.8 & 0.0 & 0.8  \\
Multilabel & N/A & True & True & False & False  \\
Loss Function & MSE & BCE & BCE & CE & BCE  \\
Dataset Mean for Normalization & -4.268 & -4.268 & -4.268 & -6.627 & -6.846  \\
Dataset Std for Normalization & 4.569 & 4.569 & 4.569 & 5.359 & 5.565  \\
\bottomrule
\end{tabular}
\caption{\textbf{Hyperparameters Configuration for EAT Pre-Training and Fine-Tuning.} }
\label{tab:training_config}
\end{table*}

\end{document}